\documentclass[aps,showpacs,prl,twocolumn]{revtex4-1}
\usepackage{graphicx}
\usepackage{dcolumn}
\usepackage{bm}

\def\be {\begin{eqnarray}}
\def\ee {\end{eqnarray}}

\begin{document}

\title{Spectroscopic evidence for two-gap superconductivity in the quasi-one dimensional chalcogenide Nb$_{2}$Pd$_{0.81}$S$_{5}$}

\author{Eunsung Park$^{1}$}
\author{Xin Lu$^{2, 3}$}
\email[]{xinluphy@zju.edu.cn}
\author{Filip Ronning$^{3}$}
\author{J. D. Thompson$^{3}$}
\author{Q. Zhang$^{4}$}
\author{L. Balicas$^{4}$}
\author{Tuson Park$^{1}$}
\email[]{tp8701@skku.edu}
\affiliation{$^{1}$Department of Physics, Sungkyunkwan University, Suwon 440-746, Korea}
\affiliation{$^2$ Center for Correlated Matter and Department of Physics, Zhejiang University, Hangzhou 310058, China}
\affiliation{$^{3}$Los Alamos National Laboratory, Los Alamos, New Mexico 87545, USA}
\affiliation{$^4$ National High Magnetic Field Laboratory, Florida State University, Tallahassee, Florida 32310, USA}


\begin{abstract}
We present the first direct evidence for two-band superconductivity in the quasi-one dimensional chalcogenide Nb$_{2}$Pd$_{0.81}$S$_{5}$. Soft point contact spectroscopic measurements reveal Andreev reflection in the differential conductance $G$ in the zero-resistance superconducting (SC) state below $T_c$(=6.6~K). Multiple peaks in $G$ were clearly observed at 1.8~K and were successfully explained by the two-band Blonder-Tinkham-Klapwijk model with two gaps $\Delta_{1}=0.48$~meV and $\Delta_{2}=1.05$~meV. Their evolution in temperature and magnetic field is consistent with the conventional BCS theory. The SC gap to $T_c$ ratio for $\Delta_1$ and $\Delta_2$ is 1.7 and 3.7, which is similar to the weak coupling BCS prediction of 3.5. These results demonstrate that the newly discovered niobium chalcogenide is a two-gap superconductor in the weak coupling limit.

\end{abstract}


\maketitle

Materials with low-dimensional electronic structure have attracted interest because they provide a rich avenue to explore various novel phases arising from strong electron correlations~\cite{gozar08, berezinskii72, kosterlitz73, dresselhaus14}. A spin (or charge) density wave is one such example that occurs due to geometry of the Fermi surface in momentum space that is conducive to the development of a Peierls instability which leads to a spatial modulation of the spins (or charges). Unconventional superconductivity with a triplet superconducting (SC) order parameter, or a chiral d-wave symmetry was also proposed due to strong repulsive electron-electron interactions~\cite{gorkov85, chubukov12}. An inhomogeneous superconducting phase with a periodic modulation in SC amplitude, or Fulde-Ferrel-Larkin-Ovchinnikov state,  may also be realized at low temperature and high magnetic fields in low-dimensional compounds~\cite{fulde64, larkin65}.

The newly discovered chalcogenide superconductor Nb$_{2}$Pd$_{x}$S$_{5}$ (or Nb215) has attracted interest because of its quasi-one-dimensional (1D) electronic structure~\cite{zhang13, niu13, singh13, zhou13, ning15, biswas15}. Spin susceptibility is estimated to be strongly enhanced at small $q$ wavevectors, indicating that Nb215 is in close proximity to a magnetically ordered state with long modulation wave-lengths. When compared to its relatively low SC transition temperature ($T_c = 6.6$~K), its upper critical field is very high ($H_{c2}=37$~T), exceeding the Pauli limiting field expected for weakly coupled superconductors ($H_p\approx  1.84T_c=12$~T). The temperature dependence of the upper critical field perpendicular to the chain direction (b-axis) does not saturate at low temperatures, but linearly increases with decreasing temperature. When combined with the strong dependence on temperature of the $H_{c2}$ anisotropy and multiple Fermi surfaces, the linear-in-$T$ dependence of $H_{c2}$ of Nb215 is indicative of multiband superconductivity~\cite{zhang13}. Recent muon spin relaxation and rotation measurements, however, claimed a single s-wave energy gap from the temperature dependence of the magnetic penetration depth, thus requiring a more definitive study on the nature of SC gap in order to understand the SC mechanism of this low-dimensional superconductor.

Here, we report the first, direct evidence for two superconducting gaps in the chalcogenide superconductor Nb$_{2}$Pd$_{x}$S$_{5}$ ($x=0.81$). Differential conductance $G (=dI/dV)$, which is obtained via soft point-contact spectroscopy, shows an enhancement at zero-bias voltage below $T_c$ due to the Andreev reflection and multiple peaks become more distinguishable with decreasing temperature. A single band BTK (Blonder-Tinkham-Klapwijk) model fails to explain the spectroscopic features deep inside the SC state, while the two band BTK model best explains the two peak structure with $\Delta_1 = 0.48$~meV and $\Delta_2=1.05$~meV at 1.8~K. The magnetic field suppresses both the large and the small SC gaps, where a least-squares-fit with the BCS model reasonably explains the field dependence of both of them with $H_{c2}=17$~T for $H//c$, which is consistent with the previous report~\cite{zhang13}. Taken together, these results prove the existence of multiple superconducting gaps in the newly discovered quasi-1D chalcogenide compound Nb$_{2}$Pd$_{x}$S$_{5}$.

\begin{figure}[tb]
\center 
\includegraphics[width=0.45\textwidth]{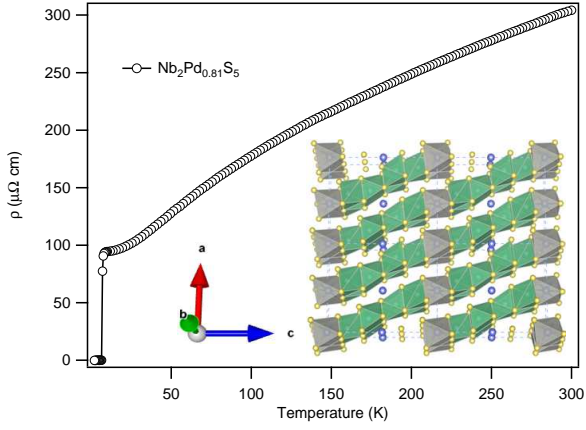}
\caption{Electrical resistance of Nb$_{2}$Pd$_{0.81}$S$_{5}$ for the electric current applied along the b-axis. Inset : Crystal structure of Nb$_{2}$Pd$_{0.81}$S$_{5}$. Yellow, blue, and green symbols represent S, Pd, and Nb atoms, respectively.}
\label{fig1}
\end{figure}

Single crystalline Nb$_{2}$Pd$_{x}$S$_{5}$ samples with $x=0.81$ were grown by solid reaction crystallizing in a monoclinic structure, where the Nb-S trigonal prisms are stacked along the b-axis forming conducting chain along that direction (see inset of Fig.~1)~\cite{keszler85, khim13}. Electrical resistivity measurements were performed through the conventional four-probe technique on a needle like single crystal, where the electrical current is applied along the elongated b-axis. As shown in Fig.~1, the resistivity ratio between room temperature and $T_c$, $\rho_b (300~K)/\rho_b (T_c)$, is approximately 3, which is similar to the one previously reported~\cite{zhang13}. The resistivity $\rho_b$ shows a sharp SC transition to a zero value at 6.6~K. In the following, we analyze the data below 6.6~K, where the zero-resistance state is established. In order to perform the soft point-contact spectroscopy (SPCS), silver grains from Dupont 4929N were applied on the surface of the sample~\cite{gonnelli10}. Magnetic field is applied perpendicularly to the needle axis, i.e., $H \perp b-$axis. Tens of contacts, made on two different crystals from the same batch, showed consistent spectroscopic signatures with respect to one another, thus ensuring reproducibility of the differential conductance data reported in this Letter. 

\begin{figure}[tb]
\center
\includegraphics[width=0.45\textwidth]{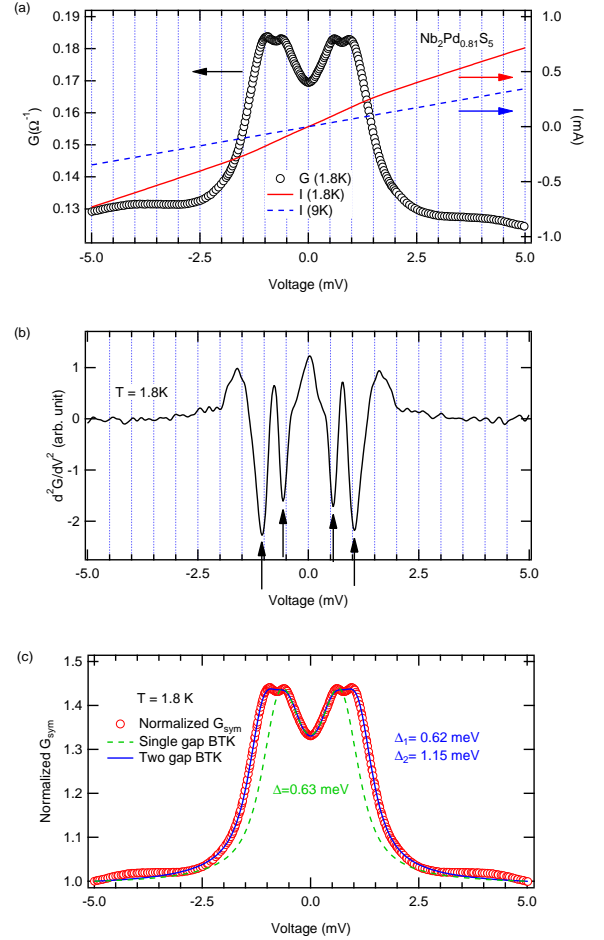}
\caption{(a) Soft point-contact spectroscopy of Nb$_{2}$Pd$_{0.81}$S$_{5}$. Differential conductance $G$ at 1.8~K is shown as a function of the bias voltage on the left ordinate (open circles). The current-voltage (I-V) characteristic curves are shown on the right ordinate for $T=$1.8~K (solid line) and 9~K (dashed line), for which the sample is in the superconducting and in the normal state, respectively. (b) The second derivative of the differential conductance, d$^2G/$d$V^2$ at 1.8~K. (c) Normalized differential conductance of the Nb215 at 1.8~K. $G_{sym}$ is the symmetrized differential conductance with respect to negative and positive bias voltages, i.e. $G_{sym}=(G_+ + G_-)/(G(V=+5mV)+G(V=-5mV))$. Here $G_+$ and $G_-$ is the differential conductance at positive and negative bias voltage, respectively. After symmetrization, $G_{sym}$ was divided by the valued at +5 mV. Dashed and solid lines describe the least-squares fits from the single band and two band BTK models, respectively.}
\label{fig2}
\end{figure}

Figure~\ref{fig2}(a) shows a representative plot of current $I$ as a function of the bias voltage $V$ (on the right ordinate) at 1.8~K (solid line) and 9~K (dashed line). The nearly linear $I-V$ characteristics for $V > \Delta$ as well as the absence of dips in the differential conductance $G$ indicate that the effective contact size is close to the Sharvin limit~\cite{PCSBOOK, gonnelli10}. Here $\Delta$ is the size of the SC gap. The dependence of $G$ on the bias voltage of $V$ at 1.8~K is shown on the left ordinate of Fig.~\ref{fig2}(a) (open circles). Multiple peaks in $G$, which are almost symmetric with respect to the zero-bias voltage, indicates two SC energy gaps for the Nb215 compound. As shown in Fig.~\ref{fig2}(b), the second derivative of the differential conductance spectra, $d^{2}G/dV^{2}$, reveals pronounced dips that correspond to the multiple peaks in $G$. Even though data are scattered, the first derivative of the $I-V$ curve at 1.8~K is consistent with the $G$ obtained from an ac modulation technique (see Fig.~S1, Supplementary Information), indicating that the multiple peak structure in $G$ represents intrinsic SC properties of the Nb215 superconductor.

In order to quantitatively understand the multiple SC gaps for Nb215, the multi-gap BTK model is used and the total conductance $G=dI/dV$ is the sum of the multigap components with a weighting factor $w_i$ for the $i$th component, or $G= \sum_{i}{\omega_{i}G_{i}(V)}$. The tunneling transparency of the point contact barrier $Z_{i}$ and quasi-particle lifetime broadening effect $\Gamma_{i}$ were parametrized in the BTK model for the conductance between normal and superconducting phases~\cite{gonnelli10, btk82, plecenik94}:
\be
\frac{dI}{dV} (V, T)= I_{0} \frac{d}{dV} \int_{-\infty}^{\infty} [f(E-eV, T) - f(E, T)] \sigma(E) dE \nonumber \\
= I_{0} \int_{-\infty}^{\infty} \frac{1}{4T} sech^2(\frac{E-eV}{2T}) \sigma(E)dE.~~~~~~~~~~~~
\ee
Here, $f(E,T)$ is the Fermi-Dirac distribution function at a finite temperature $T$ and $I_0$ is the prefactor that depends on the density of states of the metallic tip and the SC material. The BTK integral kernel, $\sigma(E)$, is written as
\be
\sigma(E)=\tau_N \frac{1+\tau_N\left|\gamma(E)\right|^2 + (\tau_N - 1)\left|\gamma(E)^2 \right|^2} {\left|1+(\tau_N -1)\gamma(E)^2\right|^2}.
\ee 
The transparency of the barrier is given by $\tau_N = 1/(1+Z^2)$ and the complex function $\gamma (E) = (N_q(E)-1)/N_p(E)$, where $N_q(E, \Gamma)=(E+i\Gamma)/\sqrt{(E+i\Gamma)^2-\Delta^2}$ and $N_p(E, \Gamma)=\Delta / \sqrt{(E+i\Gamma)^2-\Delta^2}$:

\begin{figure}[tb]
\center 
\includegraphics[width=0.45\textwidth]{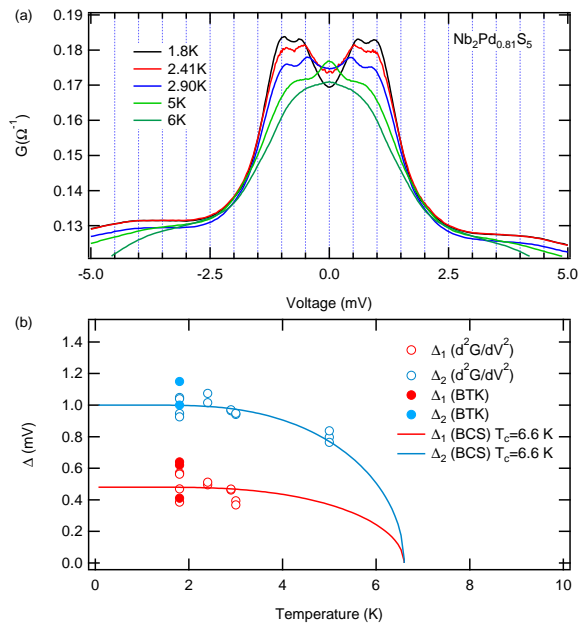}
\caption{(a) Evolution of the soft point contact Andreev reflection as a function of temperature. (b) Temperature dependence of the SC energy gaps obtained from  d$^{2}G$/dV$^{2}$. BTK results at 1.8~K are added for comparison. Solid lines are estimated from the BCS theory with $T_{c}$ of 6.6~K.}
\label{fig3}
\end{figure}
Figure~\ref{fig2}(c) shows the normalized differential conductance $G_{sym}$ at 1.8~K, where $G_+$ at positive voltage and $G_-$ at negative voltage is symmetrized and divided by the value at +5~mV, i.e., $G_{sym}=(G_+ + G_-)/(G(=+5mV)+G(=-5mV))$. A single band BTK model (dashed line) could explain the dip at zero bias voltage, but fails to explain the second dip at around $\pm 0.7$~mV. In contrast the two band BTK model (solid line) captures the characteristic features with $\Delta_1=0.62~\pm0.018$~meV, $\Delta_2=1.15~\pm0.012$~meV and the weight of the first gap $\omega_{1}=0.46$. The amplitude of the two SC gaps is close to the value of the dip positions in the second derivative of the differential conductance d$^2G/$d$V^2$, as shown in Fig.~2b. We note that a small $\Gamma $ value of less than 0.01 equally fits well the normalized conductance $G_{sym}$, indicating both a clean interface and a quasiparticle lifetime that is close to infinite at low temperatures. The least-squares fit based on the two-band BTK model best describes the data with effective tunneling barriers $Z$ of 0.3 and 0.5 for the small and large SC gaps, respectively.

\begin{figure}[tb]
\center 
\includegraphics[width=0.45\textwidth]{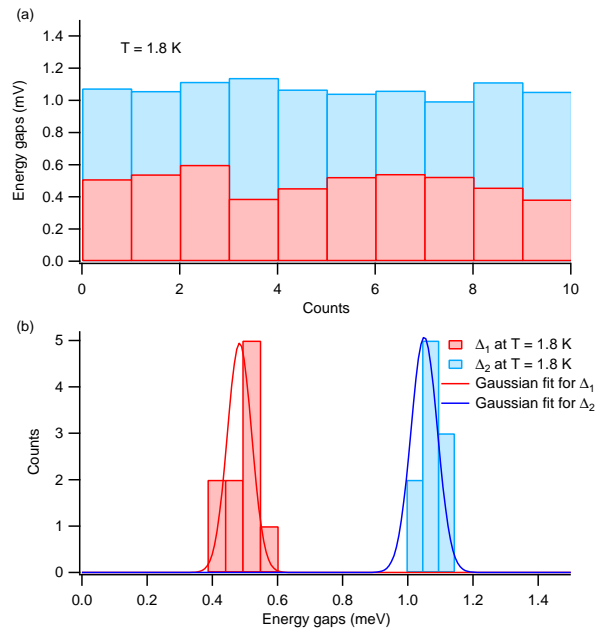}
\caption{(a) Distribution of SC energy gaps obtained from d$^{2}G$/dV$^{2}$. (b) Statistical counts of SC energy gap values for ten different contacts. Gaussian fit is used to determine the average SC gaps.}
\label{fig4}
\end{figure}

Figure~\ref{fig3}(a) selectively shows the differential conductance $G$ of Nb215 for several temperatures in its SC state. The conductance dip at the zero-bias voltage, which is typical for Andreev reflection when $Z\neq 0$, becomes shallower with increasing temperature and disappears completely for temperatures above 2.9~K due to thermal smearing effects. Another dip structure at $\pm 0.7$~mV for 1.8~K, which reflects the two-gap nature of the dichalcogenide superconductor, is also smeared out with increasing temperature due to thermal effects. The decrease in the dip position with increasing temperature is consistent with the decrease in the gap amplitude. The temperature dependence of the two SC gaps is shown in Fig.~\ref{fig3}(b), where the gap amplitudes are obtained from the two-band BTK fitting (solid symbols) and from the dip in the second derivative of the differential conductance (open symbols) for several different contacts made on the ac-plane of different crystals. Solid lines depict the dependence expected from the weak coupling BCS theory, which reasonably describes the temperature dependence of the gaps with $\Delta_1=0.48$~meV and $\Delta_2=1.05$~meV.

Figure~\ref{fig4} shows statistical counts for ten different point contacts and the distribution of SC energy gaps obtained at 1.8~K from the second derivative of the differential conductance, d$^2G/$d$V^2$, for Nb215. The small SC gap $\Delta_1$ is concentrated between 0.4 and 0.6~meV, while the large SC gap $\Delta_2$ is between 1.0 and 1.15~meV. As shown in Fig.~4(b), the statistical distribution can be explained by a Gaussian function, where the peaks are located at 0.48 $\pm$0.02 and 1.05 $\pm$0.04~meV for $\Delta_1$ and $\Delta_2$, respectively. The SC gap ratios against $T_c$, 2$\Delta_0$/k$_{B}$$T_{c}$, are 1.7 and 3.7, where the gap ratio for the large gap is close to the BCS prediction of 3.5 for weak coupling superconductors~\cite{inosov11}. Recently, the temperature dependence of the specific heat for Nb$_2$PdS$_5$ was analyzed in terms of multiple SC gaps, where the small and large gap ratio is 1.9 and 6.4, respectively~\cite{goyal15}. The difference between this work and Ref~\cite{goyal15} may arise from  the limited temperature range over which the specific heat was studied, i.e., $2$~K$< T < 6.6$~K. When measured down to 0.5~K, a similar specific heat study for another chalcogenide Nb$_2$Pd$_{1.2}$Se$_5$ showed that the gap ratios are 2.1 and 4.1, which is similar to this work~\cite{khim13}. We note that the gap values obtained from the two-band BTK model are consistent with those from d$^2G/$d$V^2$.

\begin{figure}
\center
\includegraphics[width=0.45\textwidth]{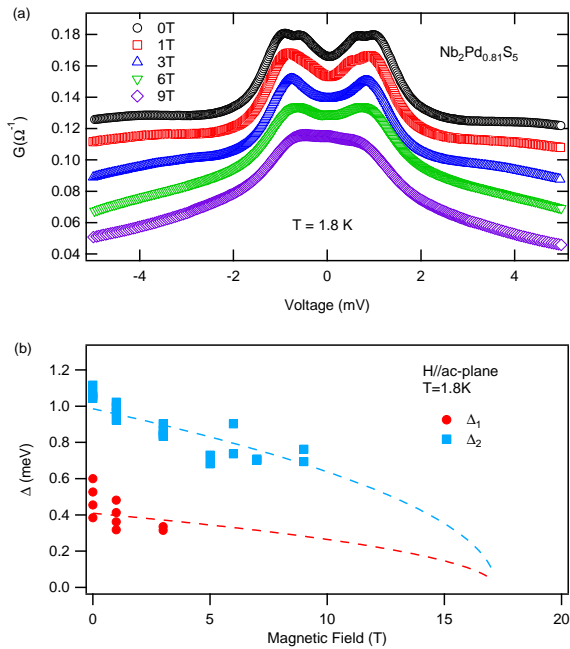}
\caption{(a) Magnetic field dependence of the differential conductance of Nb215 at 1.8~K for 0, 1, 3, 6, and 9 Tesla. The conductance for each magnetic field is vertically displaced for clarity. (b) Magnetic field dependence of the SC gap from d$^{2}G$/dV$^{2}$. Dashed lines correspond to least-squares fits from Eq~(3).}
\label{fig5}
\end{figure}
The dependence of the differential conductance of Nb215 at 1.8~K on the magnetic field $H$ is shown in Fig.~\ref{fig5}(a), where $H$ is applied perpendicularly to the crystallographic $b$-axis. For clarity, the conductance for each field is vertically displaced with respect to the zero-field spectrum. The multiple peaks observed at zero field are rapidly suppressed with increasing field, which could be due to a fast decrease in the Andreev reflection from the small SC gap $\Delta_1$ under magnetic field. The dip observed at zero-bias voltage, which appears in the Andreev reflection for a finite $Z$, becomes blurred with increasing magnetic field. The pair-breaking effects induced by the magnetic field are indicated by the enhancement in the effective broadening parameter $\Gamma$ due to a reduction in the quasiparticle lifetime~\cite{gonnelli10, naidyuk96}. 

The magnetic field dependence of the SC energy gaps as determined from the dip positions in d$^{2}G/$d$V^{2}$ is shown in Fig.~\ref{fig5}(b). The dashed lines are least-square-fits of the SC gaps for type-II superconductors, where the SC gap ratio $2\Delta / k_BT_c$ is assumed to be independent of magnetic field~\cite{morenzoni14, gross86}:
\be
\Delta(T, B)=\Delta(T, 0)\sqrt{1-B/B_{c2}(0)},
\ee
where $\Delta(T, 0)=\Delta(0,0)\text{tanh}(1.781 \sqrt{T_c(B)/T-1})$ and $T_c(B)=T_c(0)\sqrt{1-B/B_{c2}(0)}$. The upper critical field $B_{c2}$ at 1.8~K is 17~T, which is consistent with magneto-resistance measurements~\cite{zhang13}.

To conclude, we have directly shown that the quasi-1D chalcogenide Nb$_{2}$Pd$_{0.81}$S$_{5}$ is a two-gap superconductor, where the amplitude is 1.05 and 0.48~meV for the large and the small SC gaps, respectively. The differential conductance curves measured from soft point-contact spectroscopy for Nb215 were successfully explained by the two-band BTK model with a finite point contact barrier $Z$ and a quasiparticle lifetime broadening effect $\Gamma$. With increasing temperature, the multiple peak structure and the dip at the zero-bias voltage in the differential conductance become weaker due to thermal broadening. The temperature dependence as well as the magnetic field dependence of the SC gaps are in agreement with the BCS theory. The gap/temperature ratio for the larger gap is 3.7, which is slightly larger than BCS value of 3.5 for weakly coupled superconductors. Although the isotropic BTK model was successfully applied to describe  the differential conductance of the Nb215 superconductor, its quasi-1D Fermi surfaces have yet to be taken into account to make a better understanding of the point contact spectra.

Work at Sungkyunkwan University was supported by a NRF grant funded by the Ministry of Science, ICT and Future Planning (No.~2012R1A3A2048816). Work at Los Alamos was performed under the auspices of the US Department of Energy, Office of Science, Division of Materials Science. Work at Zhejiang University is supported by the National Natural Science Foundation of China (Grant No. 11374257) and the Fundamental Research Funds for the Central Universities. LB is supported by DOE-BES through award DE-SC0002613.

\end{document}